\documentclass[fleqn]{llncs}

\usepackage{latexsym}
\usepackage{amssymb}
\usepackage{amsmath}
\usepackage[all]{xy}
\newcommand{\ignore}[1]{} 
\newcommand{\too}{\longrightarrow}

\usepackage{color}

\newcommand{\blue}[1]{{\color{black} #1}}

\makeatletter
\newenvironment{prog}{\vspace{0.7ex}\par
\setlength{\parindent}{0.7cm}
\obeylines\@vobeyspaces\tt}{\vspace{0.7ex}\noindent
}
\makeatother
\newcommand{\startprog}{\begin{prog}}
\newcommand{\stopprog}{\end{prog}\noindent}

\newcommand{\id}{id}

\def\defemb#1#2{\expandafter\def\csname #1\endcsname
                              {\relax\ifmmode #2\else\hbox{$#2$}\fi}}
\defemb{cA}{{\cal A}}
\defemb{cB}{{\cal B}}
\defemb{cC}{{\cal C}}
\defemb{cD}{{\cal D}}
\defemb{cE}{{\cal E}}
\defemb{cF}{{\cal F}}
\defemb{cG}{{\cal G}}
\defemb{cH}{{\cal H}}
\defemb{cI}{{\cal I}}
\defemb{cJ}{{\cal J}}
\defemb{cL}{{\cal L}}
\defemb{cM}{{\cal M}}
\defemb{cN}{{\cal N}}
\defemb{cO}{{\cal O}}
\defemb{cP}{{\cal P}}
\defemb{cQ}{{\cal Q}}
\defemb{cR}{{\cal R}}
\defemb{cS}{{\cal S}}
\defemb{cT}{{\cal T}}
\defemb{cU}{{\cal U}}
\defemb{cV}{{\cal V}}
\defemb{cX}{{\cal X}}
\defemb{cZ}{{\cal Z}}

\newcommand{\dom}{{\cD}om}

\newcommand{\nil}{[\:]}

\renewcommand{\phi}{\varphi}

\newcommand{\ol}[1]{\overline{#1}}  

\newcommand{\cons}{\!:\!}

\def \tuple#1{\langle #1 \rangle}

\long\def\comment#1{}

\begin{document}

\title{Towards Reversible Computation in Erlang%
\thanks{This work has been partially supported by the EU (FEDER) and
  the Spanish \emph{Ministerio de Econom\'{\i}a y Competitividad}
  (MINECO) under grant TIN2013-44742-C4-1-R and by the
  \emph{Generalitat Valenciana} under grant PROMETEO-II/2015/013
  (SmartLogic). 
  The second and third authors also acknowledge a partial support
  of  COST Action IC1405 on Reversible Computation - extending
  horizons of computing.}
}

\author{Naoki Nishida\inst{1} \and Adri\'an Palacios\inst{2}%
${}^{,}$\thanks{Partially
   supported by the the EU (FEDER) and the Spanish \emph{Ayudas para
     contratos predoctorales para la formaci\'on de doctores de la
     Secretar\'{\i}a de Estado de Investigaci\'on, Desarrollo e
     Innovaci\'on del} MINECO under FPI grant BES-2014-069749.} 
  \and Germ\'an Vidal\inst{2}}

\institute{Graduate School of Information Science, Nagoya University\\ 
Furo-cho, Chikusa-ku, 4648603 Nagoya, Japan,\\
\email{nishida@is.nagoya-u.ac.jp}
\and
MiST, DSIC, Universitat Polit\`ecnica de Val\`encia\\
Camino de Vera, s/n, 46022 Valencia, Spain\\
\email{$\{$apalacios$,\>$gvidal$\}$@dsic.upv.es}}

\maketitle

\begin{abstract}
  In a reversible language, any forward computation can be undone by a
  finite sequence of backward steps. Reversible computing has been studied in the context
  of different programming languages and formalisms, where it has been
  used for debugging and for enforcing
  fault-tolerance, among others.
  In this paper, we consider a subset of Erlang, a concurrent language
  based on the actor model. We formally introduce a reversible
  semantics for this language. 
  To the best of our knowledge, this is the first attempt to define a
  reversible semantics for Erlang.
\end{abstract}

\section{Introduction} \label{intro}

Let us consider that the operational semantics of a programming
language is specified by a state transition relation $R$ such that
$R(s,s')$ holds if the state $s'$ is reachable---in one step---from
state $s$. As it is common practice, we let $R^\ast$ denote the
reflexive and transitive closure of $R$. Then, we say that a
programming language (or formalism) is \emph{reversible} if there
exists a constructive 
algorithm that can be used to, given a computation from state $s$ to
state $s'$, in symbols $R^\ast(s,s')$, obtain the state $s$ from
$s'$. In general, such a property does not hold for most programming
languages and formalisms.
We refer the interested reader to, e.g.,
\cite{Bennett00,Fra05,Yok10,YAG08} for a high level account of the
principles of reversible computation.

The notion of \emph{reversible computation} was first introduced in
Landauer's seminal work \cite{Lan61} and, then, further improved by
Bennett \cite{Ben73} in order to avoid the generation of ``garbage''
data.  The idea underlying these works is that any programming
language or formalism can be made reversible by adding the
\emph{history} of the computation to each state, which is usually
called a \emph{Landauer's embedding}. Although carrying the history of
a computation might seem infeasible because of its size, there are
several successful proposals that are based on this idea. In
particular, one can restrict the original language or apply a number
of analysis in order to restrict the required information in the
history as much as possible, as in, e.g., \cite{MHNHT07,NPV16,TA15} in
the context of a functional language.

Alternatively, one can consider a restricted language where
computations are reversible without adding any additional
information to the states. This is the case, e.g., of the functional
language considered in \cite{YAG16} or the language Janus (see
\cite{Yok10} and references therein). Obviously, such languages are
not universal since there are functions that cannot be represented
(e.g., non-injective functions).

In this paper, we consider the former approach, the so-called
Landauer's embedding. In particular, we aim at introducing a form of
reversibility in the context of a programming language that follows
the actor model (concurrency based on message passing), and that can
be considered as a subset of the concurrent and functional language
Erlang \cite{AVW96}. Previous approaches have mainly considered
reversibility in---mostly synchronous---concurrent calculi like CCS
\cite{DK04,DK05}, a general framework for reversibility of algebraic
process calculi \cite{PU07}, or the recent approach to reversible
\emph{session-based} $\pi$-calculus \cite{TY15}. However, we can only
find a few approaches that considered the reversibility of
\emph{asynchronous} calculi, e.g., Cardelli and Laneve's reversible
structures \cite{CL11} or the approach based on a rollback construct
of \cite{GLM14,LMSS11,LMS16,LLMS12} for a higher-order asynchronous
$\pi$-calculus.

To the best of our knowledge, our work is the first one that considers
reversibility in the context of the functional, concurrent, and
distributed language Erlang. Here, given a running Erlang system
consisting of a pool of interacting processes, possibly distributed in
several computers, we aim at allowing a \emph{single} process to undo
its actions, including the interactions with other processes,
following a rollback fashion. In this context, we must ensure
\emph{causal consistency} \cite{DK04}, i.e., an action cannot be
undone until all the actions that depend on it have been already
undone. E.g., if a process spawns another process, we cannot undo this
process spawning until all the actions performed by the new process
are undone too. This is particularly challenging in our asynchronous
and distributed setting since there is no \emph{global} order for the
language events. In this paper, we introduce a rollback operator that
undoes the actions of a process until a \emph{checkpoint} is
reached. This could be considered as a promising basis for defining
\emph{safe} sessions in a language like Erlang.

In this paper, we consider a simple Erlang-like language that can be
considered a subset of \emph{Core Erlang} \cite{CGJLNPV04}.
We present the following contributions:
First, we introduce an appropriate standard semantics for the
language. In contrast to monolothic previous semantics like that in
\cite{CMRT13tr}, our semantics is more modular, which simplifies the
definition of a reversible extension. In contrast to \cite{SFB10},
although we follow some of the ideas in this approach (e.g., the use
of a global mailbox), we include the evaluation of expressions and,
moreover, our treatment of messages is more
deterministic.\footnote{E.g., in the semantics of \cite{SFB10}, at the
  expression level, the transition semantics of an expression
  containing a receive statement is, in principle, infinitely
  branching, since their formulation allows for an infinite number of
  possible queues and selected messages.}
We then introduce a reversible extension of the standard semantics
(basically, a Landauer's embedding). Here, we focus only on the
concurrent actions (namely, process spawning, message sending and
receiving) and, thus, do not consider the reversibilization of the
functional component of the language; rather, we assume that the state
of the process---the current expression and its environment---is
stored in the history after each execution step. This approach could
be improved following, e.g., the approaches from
\cite{MHNHT07,NPV16,TA15}.
Finally, we introduce a backward semantics that can be used to undo
the actions of a given process in a rollback fashion, until a
checkpoint---introduced by the programmer---is reached. Here, ensuring
causal consistency is essential and might propagate the rollback
action to other, dependent processes.

\section{Language Syntax} \label{syntax-sec} 

In this section, we present the syntax of a first-order concurrent and
distributed functional language that follows the actor model. Our
language is basically equivalent to a subset of Core Erlang
\cite{CGJLNPV04}.

\begin{figure}[t]
  \begin{center}
  $
  \begin{array}{rcl@{~~~~~~}l}
    \mathit{Module} & ::= & \mathsf{module} ~ Atom = 
    \mathit{fun}_1,\ldots,\mathit{fun}_n\\
    {\mathit{fun}} & ::= & \mathit{fname} = \mathsf{fun}~(X_1,\ldots,X_n) \to expr \\
    {\mathit{fname}} & ::= & Atom/Integer \\
    lit & ::= & Atom \mid Integer \mid Float \mid \nil \\
    expr & ::= & \mathit{Var} \mid lit \mid \mathit{fname} \mid [expr_1|expr_2]
                 \mid   \{expr_1,\ldots,expr_n\} \\
    & \mid & \mathsf{call}~expr~(expr_1,\ldots,expr_n) 
    \mid \mathsf{apply}~expr~(expr_1,\ldots,expr_n) \\
    & \mid &
    \mathsf{case}~expr~\mathsf{of}~clause_1;\ldots;clause_m~\mathsf{end}\\
    & \mid & \mathsf{let}~\mathit{Var}=expr_1~\mathsf{in}~expr_2 
    \mid \mathsf{receive}~clause_1;\ldots;clause_n~\mathsf{end}\\
    & \mid & \mathsf{spawn}(expr,[expr_1,\ldots,expr_n])  
     \mid expr_1 \:!\: expr_2 \mid \mathsf{self}()\\
    clause & ::= & pat ~\mathsf{when}~expr_1 \to expr_2
    \\
    pat & ::= & \mathit{Var} \mid lit \mid [pat_1|pat_2] \mid
    \{pat_1,\ldots,pat_n\} \\
  \end{array}
  $
  \end{center}
\vspace{-2ex}
\caption{Language syntax rules} \label{ErlangSyntax}
\vspace{-2ex}
\end{figure}

The syntax of the language can be found in Figure~\ref{ErlangSyntax}.
Here, a module is a sequence of function definitions, where each
function name $f/n$ (atom/arity) has an associated definition of the
form $\mathsf{fun}~(X_1,\ldots,X_n) \to e$. We consider that a program
consists of a single module for simplicity. The body of a function is
an \emph{expression}, which can include variables, literals, function names,
lists, tuples, calls to built-in functions---mainly arithmetic and
relational operators---, function applications, case expressions, let
bindings, and receive expressions; furthermore, we also consider the
functions $\mathsf{spawn}$, ``$!$'' (for sending a message), and
$\mathsf{self}()$ that are usually considered built-in's in the Erlang
language.

Despite the general syntax in Figure~\ref{ErlangSyntax}, as mentioned
before, we only consider first order expressions. Therefore, the first
expression in calls, applications and spawns can only be function
names (instead of arbitrary expressions or closures).

In this language, we distinguish expressions, patterns, and values. As
mentioned before, expressions can include all constructs of the
language. In contrast, \emph{patterns} are built from variables,
literals, lists, and tuples. Finally, \emph{values} are built from
literals, lists, and tuples, i.e., they are \emph{ground}---without
variables---patterns. Expressions are denoted by
$e,e',e_1,e_2,\ldots$, patterns by $p,p',p_1,p_2,\ldots$ and values by
$v,v',v_1,v_2,\ldots$
As it is common practice, a \emph{substitution} $\theta$ is a mapping
from variables to expressions such that $\dom(\theta) =
\{X\in\mathit{Var} \mid X \neq \theta(X)\}$ is its
domain. Substitutions are usually denoted by sets of mappings like,
e.g., $\{X_1\mapsto v_1,\ldots,X_n\mapsto v_n\}$.
Substitutions are extended to morphisms from expressions to
expressions in the natural way.  
The identity substitution is denoted by $\id$. Composition of
substitutions is denoted by juxtaposition, i.e., $\theta\theta'$
denotes a substitution $\theta''$ such that $\theta''(X) =
\theta'(\theta(X))$ for all $X\in\mathit{Var}$. Also, we denote by
$\theta[X_1\mapsto v_1,\ldots,X_n\mapsto v_n]$ the \emph{update} of
$\theta$ with the mapping $X_1\mapsto v_1,\ldots,X_n\mapsto v_n$,
i.e., it denotes a new substitution $\theta'$ such that $\theta'(X) =
v_i$ if $X = X_i$, for some $i\in\{1,\ldots,n\}$, and $\theta'(X) =
\theta(X)$ otherwise. 

In a case expression 
``$
\mathsf{case}~e~\mathsf{of} ~ p_1~\mathsf{when}~e_1 \to
e'_1; ~
   \ldots; ~ 
   p_n~\mathsf{when}~e_n \to e'_n~~\mathsf{end} 
$''\!\!,
we first evaluate
$e$ to a value, say $v$; then, we should find (if any) the first
clause $p_i ~\mathsf{when}~e_i\to e'_i$ such that $v$ matches
$p_i$ (i.e., there exists a substitution $\sigma$ for the variables
of $p_i$ such that $v=p_i\sigma$) and $e_i\sigma$---the
\emph{guard}---reduces to $\emph{true}$; then, the
case expression reduces to $e'_i\sigma$. Note that guards can only
contain calls to built-in functions (typically, arithmetic and
relational operators).

As for the concurrent features of the language, we consider that a
\emph{system} is of a pool of processes that can only interact through
message sending and receiving (i.e., there is no shared memory). Each
process has an associated \emph{pid} (process identifier), which is
unique in a system. For clarity, we often denote pids with roman
letters $\mathrm{p},\mathrm{p'},\mathrm{p_1},\ldots$, though they are
considered values in our language (i.e., atoms). By abuse of notation,
when no confusion can arise, we refer to a process with its pid. 

An expression of the form $\mathsf{spawn}(f/n,[e_1,\ldots,e_n])$ has,
as a \emph{side effect}, the creation of a new process with a fresh
pid $\mathrm{p}$ which is initialized with the expression
$\mathsf{apply}~f/n~(e_1,\ldots,e_n)$; the expression
$\mathsf{spawn}(f/n,[e_1,\ldots,e_n])$ itself evaluates to the new pid
$\mathrm{p}$.
The function $\mathsf{self}()$ just returns the pid of the current
process.  
An expression of the form $\mathrm{p}\:!\: v$ evaluates to the value
$v$ and, as a side effect, stores the value $v$---the
\emph{message}---in the queue or \emph{mailbox} of process
$\mathrm{p}$.

Finally, an expression
``$
\mathsf{receive}~p_1~\mathsf{when}~e_1\to
e'_1;\ldots;p_n~\mathsf{when}~e_n\to e'_n~~\mathsf{end} 
$''
traverses the messages in the process' queue until one of them
matches a branch in the receive statement; i.e., it should
find the \emph{first} message $v$ in the process' queue (if any) such
that $\mathsf{case}~v~\mathsf{of}~p_1~\mathsf{when}~e_1\to
e'_1;\ldots;p_n~\mathsf{when}~e_n\to e'_n~\mathsf{end}$ can be reduced; then, the
receive expression evaluates to the same expression to which the above
case expression would be evaluated, with the additional side effect of
deleting the message $v$ from the process' queue.
If there is no matching message in the queue, the process suspends its
execution until a matching message arrives.

\begin{example} \label{ex1} Consider the program shown in
  Figure~\ref{fig-ex1}, where the symbol ``$\_$'' is used to denote an
  \emph{anonymous} variable, i.e., a variable whose name is not
  relevant. The computation starts with
  ``$ \mathsf{apply}~main/0~() $.\!''  Then, this process, say $P1$, spawns
  two new processes, say $P2$ and $P3$, and then sends the message
  ``$world$'' to process $P3$ and the message $\{P3,hello\}$ to
  process $P2$, which then resends ``$hello$'' to $P3$.
  In our language, 
  there is no guarantee regarding which message arrives first to $P3$,
  i.e., ``$\mathsf{apply}~main/0~()$'' can evaluate
  nondeterministically to either $\{hello,world\}$ or
  $\{world,hello\}$. This is coherent with the semantics of Erlang,
  where the only guarantee is that if two messages are sent from
  process $\mathrm{p}$ to process $\mathrm{p'}$ and both are
  delivered, then the order of these messages is kept.

  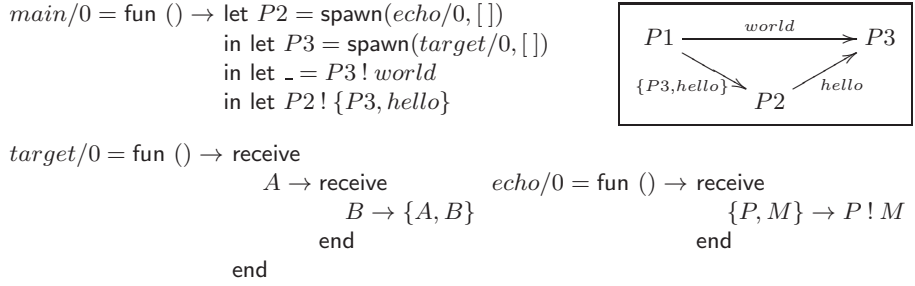
\begin{figure}[t]
    \begin{minipage}{.6\linewidth}
    $
    \begin{array}{r@{~}l@{~~}r@{~}l}
      main/0 = \mathsf{fun}~()\to & \mathsf{let}~P2 =
      \mathsf{spawn}(echo/0,\nil) \\
      & \mathsf{in}~\mathsf{let}~P3 = \mathsf{spawn}(target/0,\nil) \\
      & \mathsf{in}~\mathsf{let}~\_ = P3\:!\: world \\
      & \mathsf{in}~\mathsf{let}~P2\:!\:\{P3,hello\}\\[1ex]
     \end{array}
     $
     \end{minipage} 
     \hspace{5ex}
     \begin{minipage}{.35\linewidth}
       $ \fbox{\xymatrix@R=12pt{
         P1 \ar[rd]_{\{P3,hello\}} \ar[rr]^{world} & & P3 \\
         & P2 \ar[ur]_{hello} & \\
       }}
       $
     \end{minipage}\\[1ex]
     \begin{minipage}{\linewidth}
     $
     \begin{array}{r@{~}l@{~~}r@{~}l}
      target/0 = \mathsf{fun}~()\to & \mathsf{receive}\\

      & \hspace{3ex}A \to \mathsf{receive}~
      & echo/0 = \mathsf{fun}~()\to & \mathsf{receive}\\

      & \hspace{11ex}B\to \{A,B\}
      & & \hspace{3ex}\{P,M\}\to P\:!\: M~\\

      & \hspace{8.5ex}\mathsf{end}& &\mathsf{end} \\
      &\mathsf{end}
    \end{array}
    $
    \end{minipage}
  \caption{A simple concurrent program} \label{fig-ex1}
\vspace{-2ex}
  \end{figure}
\end{example}

\section{The Language Semantics} \label{semantics-sec} 

In order to set precisely the framework for our proposal, 
in this section we formalize the semantics of the considered
language. 

\begin{definition}[process]
  A process is denoted by a tuple of the form
  $\tuple{\mathrm{p},(\theta,e),q}$ where $\mathrm{p}$ is the pid of
  the process, $(\theta,e)$ is the control of the state---which
  consists of an environment (a substitution) and an expression to be
  evaluated---, and $q$ is the process' mailbox, a FIFO queue with the
  sequence of messages that have been sent to the process.
\end{definition}
A running \emph{system} is then a pool of processes, which is formally
defined as follows:

\begin{definition}[system]
  A system is denoted by $\Gamma;\Pi$, where $\Gamma$ is a global
  mailbox of the system (see below) and $\Pi$ is a pool of processes,
  denoted by an expression of the form
  $\tuple{\mathrm{p_1},(\theta_1,e_1),q_1} ~\& \cdots
  \&~\tuple{\mathrm{p_n},(\theta_n,e_n),q_n}$, where ``$\&$'' is an
  associative and commutative operator.
  We typically denote a system by an expression of the form 
  $
  \Gamma; \tuple{\mathrm{p},(\theta,e),q}\&\Pi
  $
  to point out that $\tuple{\mathrm{p},(\theta,e),q}$ is an arbitrary
  process of the pool (thanks to the fact that ``$\&$'' is associative
  and commutative).
\end{definition}
The role of $\Gamma$ (which is similar to the ``ether'' in
\cite{SFB10}) will be clarified later, but it is essential to
guarantee that all admissible interleavings can be modelled with the
semantics. Here, we define $\Gamma$ as a set of FIFO queues among all
(non-necessarily different) pids, i.e., $\Gamma$ is made of elements
of the form $(\mathrm{p},\mathrm{q},[v_1,\ldots,v_n])$, where
$\mathrm{p},\mathrm{q}$ are (not necessarily different) pids and
$[v_1,\ldots,v_n]$ is a (possibly empty) ordered list of messages such
that $v_1$ is the oldest message and $v_n$ is the most recent one. For
simplicity, we assume that $\Gamma$ is initialized as follows: $\{
(\mathrm{p},\mathrm{q},\nil) \mid \mathrm{p},\mathrm{q}~\mbox{are
  pids}\}$. Then, we use the following notation: $\Gamma\cup
(\mathrm{p},\mathrm{q},v)$ denotes
$\Gamma\setminus\{(\mathrm{p},\mathrm{q},vs)\} \cup
\{(\mathrm{p},\mathrm{q},vs\mathtt{+\!+}[v]\}$, while $\Gamma\setminus
(\mathrm{p},\mathrm{q},v)$ denotes
$\Gamma\setminus\{(\mathrm{p},\mathrm{q},v:vs)\} \cup
\{(\mathrm{p},\mathrm{q},vs)\}$, where $\mathtt{+\!+}$ is the list
concatenation operator.

In the following, we denote by $\ol{o_n}$ the sequence of syntactic
objects $o_1,\ldots,o_n$ for some $n$. We also write $\ol{o_{i,j}}$
for the sequence $o_i,\ldots,o_j$ when $i\leq j$ (and the empty
sequence otherwise).  We write $\ol{o}$ when the number of elements is
not relevant.

\begin{figure}[t]
  \centering
  $
  \begin{array}{c}
    (\mathit{Var}) ~ {\displaystyle \frac{}{\theta,X
        \stackrel{\tau}{\too} \theta,\theta(X)}} 


    \hspace{2ex}

    (\mathit{Tuple}) ~ {\displaystyle 
      \frac{\theta,e_i \stackrel{\ell}{\too}
        \theta',e'_i~~~i\in\{1,\ldots,n\}}{\theta,\{e_1,\ldots,e_n\}
        \stackrel{\ell}{\too} \theta',
        \{\ol{e_{1,{i-1}}},e'_i,\ol{e_{{i+1},n}}\}}}\\[4ex] 

    (\mathit{List1})  ~{\displaystyle 
      \frac{\theta,e_1 \stackrel{\ell}{\too}
        \theta',e'_1}{\theta,[e_1|e_2]
        \stackrel{\ell}{\too} \theta',
        [e'_1|e_2]}} 

    \hspace{2ex}

    (\mathit{List2}) ~ {\displaystyle 
      \frac{\theta,e_2 \stackrel{\ell}{\too}
        \theta',e'_2}{\theta,[e_1|e_2]
        \stackrel{\ell}{\too} \theta',
        [e_1|e'_2]}} \\[4ex]

      (\mathit{Let1}) ~ {\displaystyle \frac{\theta,e_1
          \stackrel{\ell}{\too} \theta',e'_1 }{\theta,\mathsf{let}~ 
          X=e_1~\mathsf{in}~e_2
          \stackrel{\ell}{\too} \theta',\mathsf{let}~ 
          X=e'_1~\mathsf{in}~e_2}}

      \hspace{1ex}

      (\mathit{Let2}) ~ {\displaystyle \frac{}{\theta,\mathsf{let}~ 
          X=v~\mathsf{in}~e
          \stackrel{\tau}{\too} \theta[X\mapsto v],e}} \\[4ex]
 
     (\mathit{Case1}) ~ {\displaystyle
        \frac{\theta,e\stackrel{\ell}{\too}
          \theta',e'}{\begin{array}{l}
            \theta,\mathsf{case}~e~\mathsf{of}~cl_1;\ldots;cl_n~\mathsf{end}\\
        \stackrel{\ell}{\too}
        \theta',\mathsf{case}~e'~\mathsf{of}~cl_1;\ldots;cl_n~\mathsf{end}\\
      \end{array}}}

  \hspace{1ex}
  
      (\mathit{Case2}) ~ {\displaystyle
        \frac{\mathsf{match}(v,cl_1,\ldots,cl_n) = \tuple{\theta_i,e_i}}{\theta,\mathsf{case}~v~\mathsf{of}~cl_1;\ldots;cl_n~\mathsf{end}
        \stackrel{\tau}{\too} \theta\theta_i,e_i}} \\[7ex]

    (\mathit{Call1}) ~ {\displaystyle
    \frac{\theta,e_i\stackrel{\ell}{\too}
    \theta',e'_i~~~i\in\{1,\ldots,n\}}{\theta,\mathsf{call}~op~(\ol{e_n})
    \stackrel{\ell}{\too}
    \theta',\mathsf{call}~op~(\ol{e_{1,i-1}},e'_i,\ol{e_{i+1,n}})}} 

    \hspace{1ex}

    (\mathit{Call2}) ~ {\displaystyle
      \frac{\mathsf{eval}(op,v_1,\ldots,v_n)=v}{\theta,\mathsf{call}~op~(v_1,\ldots,v_n)
        \stackrel{\tau}{\too} \theta,v}} \\[4ex]

    (\mathit{Apply1}) ~ {\displaystyle
      \frac{\theta,e_i\stackrel{\ell}{\too}
        \theta',e'_i~~~i\in\{1,\ldots,n\}}{\theta,\mathsf{apply}~a/n~(\ol{e_n}) 
        \stackrel{\ell}{\too}
        \theta',\mathsf{apply}~a/n~(\ol{e_{1,i-1}},e'_i,\ol{e_{i+1,n}})}}\\[5ex] 

    (\mathit{Apply2}) ~ {\displaystyle
      \frac{\mu(a/n) = \mathsf{fun}~(X_1,\ldots,X_n)\to e}{\theta,\mathsf{apply}~a/n~(v_1,\ldots,v_n)
        \stackrel{\tau}{\too} \{X_1\mapsto v_1,\ldots,X_n\mapsto v_n\},e}} 
  \end{array}
  $
\caption{Standard semantics: evaluation of sequential expressions} \label{fig:seq-rules}
\vspace{-2ex}
\end{figure}

\begin{figure}[t]
 \centering
  $
  \begin{array}{r@{~}c}
      (\mathit{Send1}) & {\displaystyle 
      \frac{\theta,e_1 \stackrel{\ell}{\too} \theta',e'_1}{\theta,e_1\:!\: e_2 \stackrel{\ell}{\too}
        \theta',e'_1\:!\: e_2} ~~~~ \frac{\theta,e_2 \stackrel{\ell}{\too} \theta',e'_2}{\theta,e_1\:!\: e_2 \stackrel{\ell}{\too}
        \theta',e_1\:!\: e'_2} 
      }\\[4ex]

      (\mathit{Send2}) & {\displaystyle
          \frac{}{\theta,v_1\:!\: v_2 \stackrel{\mathsf{send}(v_1,v_2)}{\too} \theta,v_2}
          }\\[4ex]

    (\mathit{Receive}) & {\displaystyle
      \frac{}{\theta,\mathsf{receive}~cl_1;\ldots;cl_n~\mathsf{end}
        \stackrel{\mathsf{rec}(y,\ol{cl_n})}{\too}
        \theta,y
      }
      }\\[4ex]

     (\mathit{Spawn}) & {\displaystyle
       \frac{}{\theta,\mathsf{spawn}(a/n,[e_1,\ldots,e_n])
         \stackrel{\mathsf{spawn}(y,a/n,[\ol{e_n}])}{\too} \theta,y         
       }}\\[4ex]

    (\mathit{Self}) & {\displaystyle
     \frac{}{\theta,\mathsf{self}() \stackrel{\mathsf{self}(y)}{\too} \theta,y}}
  \end{array}
  $
\caption{Standard semantics: evaluation of concurrent expressions} \label{fig:concurrent-rules}
\vspace{-2ex}
\end{figure}

The semantics is defined by means of two transition relations: $\too$
for expressions and $\longmapsto$ for systems. Let us first consider
the labelled transition relation 
\[
\too\; : (Env,Exp)\times Label\times(Env,Exp)
\]
where $Env$ and $Exp$ are the domains of environments (i.e.,
substitutions) and expressions, respectively, and $Label$ denotes an
element of the set 
\[
\{\tau, \mathsf{send}(v_1,v_2),
\mathsf{rec}(y,\ol{cl_n}), \mathsf{spawn}(y,a/n,[\ol{e_n}]),
\mathsf{self}(y)\}
\]
whose meaning will be explained below.
For clarity, we divide the transition rules of the semantics for
expressions in two sets, depicted in Figures~\ref{fig:seq-rules} and
\ref{fig:concurrent-rules} for sequential and concurrent expressions,
respectively.  

Most of the rules are self-explanatory. In the following, we only
discuss some subtle or complex issues. In principle, the transitions
are labelled either with $\tau$ (a sequential expression) or with a
label that identifies a concurrent action. Labels are used in the
system rules (Figure~\ref{fig:system-rules}) to perform the associated
side effects. 

In some of the rules (e.g., for evaluating tuples, lists, etc) we
consider for simplicity that elements are evaluated in a
non-deterministic way. In an actual programming language the order of
evaluation of the arguments in the tuple or list is usually
fixed. E.g., in Erlang, reduction takes place from left to right.

For case evaluation, we assume an auxiliary function $\mathsf{match}$
which selects the first clause, $cl_i = (p_i~\mathsf{when}~e'_i\to
e_i)$, such that $v$ matches $p_i$, i.e., $v=\theta_i(p_i)$, and the
guard holds, i.e., 
$
\theta\theta_i,e'_i \too^\ast \theta',true
$. Note that, for simplicity, we do not consider here the case in which
the argument $v$ matches no clause.

Function calls can either be defined in the program ($\mathsf{apply}$)
or be a built-in ($\mathsf{call}$). In the latter case, they are
evaluated using the auxiliary function $\mathsf{eval}$. 
In rule $\mathit{Apply2}$, we consider that the mapping $\mu$ stores
all function definitions, i.e., it maps every function name $a/n$ to
its definition $\mathsf{fun}~(X_1,\ldots,X_n)\to e$ in the program.
As for the applications, note that we only consider first-order
functions. In order to extend our semantics to also consider
higher-order functions, one should reduce the function name to a
\emph{closure} of the form $(\theta',\mathsf{fun}~(X_1,\ldots,X_n)\to
e)$ and, then, reduce $e$ in the environment $\theta'[X_1\mapsto
v_1,\ldots,X_n\mapsto v_n]$. We skip this extension since it is
orthogonal to our approach.

Let us now consider the evaluation of concurrent expressions that
produce some side effect (Figure~\ref{fig:concurrent-rules}). Here, we
can distinguish two kinds of rules. On the one hand, we have the rules
for ``$!$'', $\mathit{Send1}$ and $\mathit{Send2}$. In this case, we
know \emph{locally} what the expression should be reduced to (i.e.,
$v_2$ in rule $\mathit{Send2}$). For the remaining rules, this is not
known locally and, thus, we return a fresh distinguished symbol,
$y\not\in\mathit{Var}$ (by abuse, $y$ is dealt with as a variable), so
that the system rules will eventually bind $y$ to its correct
value. This \emph{trick} allows us to keep the rules for expressions
and systems separated (i.e., the semantics shown in
Figures~\ref{fig:seq-rules} and \ref{fig:concurrent-rules} is mostly
independent from the rules in Figure~\ref{fig:system-rules}), in
contrast to other calculi, e.g., \cite{CMRT13tr}, where they are
combined into a single transition relation.

\begin{figure}[t]
 \centering
  $
  \begin{array}{r@{~~}c}
    (\mathit{Exp}) & {\displaystyle
      \frac{\theta,e\stackrel{\tau}{\to} \theta',e'
      }{\Gamma;\tuple{\mathrm{p},(\theta,e),q}\& \Pi \longmapsto
      \Gamma;\tuple{\mathrm{p},(\theta',e'),q}\& \Pi}
      }\\[4ex]

    (\mathit{Send}) & {\displaystyle
      \frac{\theta,e \stackrel{\mathsf{send}(\mathrm{p''},v)}{\to}
        \theta',e'}{\Gamma;\tuple{\mathrm{p},(\theta,e),q} 
        \& \Pi \longmapsto \Gamma\cup (\mathrm{p},\mathrm{p''},v);\tuple{\mathrm{p},(\theta',e'),q}\& \Pi}
      }\\[4ex]

      (\mathit{Receive}) & {\displaystyle
        \frac{\theta,e \stackrel{\mathsf{rec}(y,\ol{cl_n})}{\to}
          \theta',e'~~~ \mathsf{matchrec}(\ol{cl_n},q) =
         (\theta_i,e_i,q')}
{\Gamma;\tuple{\mathrm{p},(\theta,e),q}\& \Pi \longmapsto
          \Gamma;\tuple{\mathrm{p},(\theta'\theta_i,e'\{y\mapsto e_i\}),q'}\& \Pi}
      }\\[4ex]
      
      (\mathit{Spawn}) & {\displaystyle
        \frac{\theta,e \stackrel{\mathsf{spawn}(y,a/n,[\ol{e_n}])}{\to}
          \theta',e'~~~ \mathrm{p'}~\mbox{is a fresh pid}}
{\Gamma;\tuple{\mathrm{p},(\theta,e),q} 
          \& \Pi \longmapsto \Gamma;\tuple{\mathrm{p},(\theta',e'\{y\mapsto \mathrm{p'}\}),q}\& \tuple{\mathrm{p'},(\theta',\mathsf{apply}~a/n~(e_1,\ldots,e_n)),\nil} 
          \& \Pi}
      }\\[4ex]

    (\mathit{Self}) & {\displaystyle
      \frac{\theta,e \stackrel{\mathsf{self}(y)}{\to} \theta',e'}
{\Gamma;\tuple{\mathrm{p},(\theta,e),q} 
        \& \Pi \longmapsto \Gamma;\tuple{\mathrm{p},(\theta',e'\{y\mapsto \mathrm{p}\}),q} 
        \& \Pi }
      }\\[4ex]

    (\mathit{Sched}) & {\displaystyle
      \frac{\alpha(\Gamma)=(\mathrm{p'},\mathrm{p})~~~~\Pi=\tuple{\mathrm{p},(\theta,e),q}\&\Pi'}{\Gamma;\Pi 
          \longmapsto \Gamma\setminus\{(\mathrm{p'},\mathrm{p},v)\};\tuple{\mathrm{p},(\theta,e),v\cons q}\&\Pi' }
      }

  \end{array}
  $
\caption{Standard semantics: system rules} \label{fig:system-rules}
\vspace{-2ex}
\end{figure}

Let us finally consider the system rules, which are depicted in
Figure~\ref{fig:system-rules}. In most of the rules, we consider an
arbitrary system of the form
$
\Gamma;\tuple{\mathrm{p},(\theta,e),q}\& \Pi
$,
where $\Gamma$ is the global mailbox and
$\tuple{\mathrm{p},(\theta,e),q}\& \Pi$ is a pool of process that
contains at least one process $\tuple{\mathrm{p},(\theta,e),q}$.

Note that, in rule $\mathit{Send}$, we add the triple
$(\mathrm{p},\mathrm{p''},v)$ to $\Gamma$ instead of adding it to the
queue of process $\mathrm{p''}$. This is necessary to ensure that all
possible non-deterministic results can be obtained (as discussed in
Example~\ref{ex1}). 
Observe that $e'$ is usually different from
$v$ since $e$ may have different nested operators. E.g., if $e$ has
the form ``$\mathsf{case}~\mathrm{p}\:!\:
v~\mathsf{of}~\{\ldots\}$,\!'' then $e'$ will be ``$\mathsf{case}~
v~\mathsf{of}~\{\ldots\}$'' with label $\mathsf{send}(\mathrm{p},v)$.

In rule $\mathit{Receive}$, we use the auxiliary function
$\mathsf{matchrec}$ to evaluate a receive expression. The main
difference with $\mathsf{match}$ is that $\mathsf{matchrec}$ also
takes a queue $q$ and returns the modified queue $q'$. Then, the
distinguished variable $y$ is bound to the expression in the selected
clause, $e_i$, and the environment is extended with the matching
substitution.
If no message in the queue $q$ matches any clause, then the rule is
not applicable and the selected process cannot be reduced (i.e., it
suspends).

With the rules presented so far, any system will soon reach a state in
which no reduction can be performed, since messages are stored in the
global mailbox, but they are not dispatched to the queues of the
processes. This is precisely the task of the scheduler, which is
modelled by rule $\mathit{Sched}$. The rule is non-deterministic, so
any scheduling policy can be modelled by the semantics. A message is
selected from the list of messages by the auxiliary function $\alpha$,
which can select any arbitrary pair of (non-necessarily different)
pids $(\mathrm{p'},\mathrm{p})$.
Note that we take the oldest message in the queue---the first one in
the list---, which is necessary to ensure that ``the messages
sent---directly---between two given processes arrive in the same order
they were sent'', as mentioned in the previous section.

\begin{example} \label{ex2} Let us consider the program shown in
  Figure~\ref{fig:ex2-program} and a possible execution trace. This
  trace is modelled by our semantics. For clarity, we only show in
  Figure~\ref{fig:ex2-derivation} the transition steps that correspond
  to the last two messages sent between $\mathit{client1}$ and
  $\mathit{server}$.

  \begin{figure}[t]
    \scriptsize
    \centering
      \begin{minipage}{.45\linewidth}\hspace{-8ex}
        $
        \begin{array}{r@{~}ll}
        main/0 = \mathsf{fun}~()\to & \mathsf{let}~S =
        \mathsf{spawn}(server/0,\nil)\\
        & \mathsf{in}~\mathsf{let}~\_ = \mathsf{spawn}(client/1,[S]) \\
        & \mathsf{in}~\mathsf{apply}~client/1~(S)\\[1ex]

        server/0 = \mathsf{fun}~()\to & \mathsf{receive}\\
        &\hspace{0ex}\{P,M\}\to \mathsf{let}~\_=P\:!\:
        ack~\\
        &\hspace{11ex}\mathsf{in}~\mathsf{apply}~server/0~()\\
        &\mathsf{end}\\[1ex]

        client/1 = \mathsf{fun}~(S)\to & 
        \mathsf{let}~\_=S\:!\:\{\mathsf{self}(),req\} \\
        & \mathsf{in}~\mathsf{receive}\\
        &\hspace{3ex} \mathsf{ack}\to \mathsf{ok}\\
        & \mathsf{end} \\
      \end{array}
      $
      \end{minipage}
      \hspace{5ex}
      \begin{minipage}{.4\linewidth}
      $
      \xymatrix@R=8pt@C=2pt{
        main/0 \ar@{=}[d] & & \\
        \underline{client1} \ar@{..}[ddd] \ar@/^/[r]_{\mathsf{spawn}}
        \ar@/^{8mm}/[rr]_{\mathsf{spawn}} 
        & \underline{server} \ar@{..}[d] &
        \underline{client2} \ar@{..}[d] \\
        & \mathsf{receive}
        \ar@{..}[d] &  server\:!\:req \ar@{..}[d]
        \ar[l]\\
        & client2\:!\: ack \ar@{..}[d]
       \ar[r] & \mathsf{receive}
        \ar@{..}[d] \\
        \blue{server\:!\: req} \ar@{..}[d] \ar@[blue][r] & \blue{\mathsf{receive}}
        \ar@{..}[d] &  ok \\
        \blue{\mathsf{receive}} \ar@{..}[d]  &  \ar@[blue][l]  \blue{client1\:!\:ack} \ar@{..}[d] & \\
        \blue{ok} & \blue{\mathsf{receive}} &  \\
      } 
      $
    \end{minipage}
    \caption{A simple client-server} \label{fig:ex2-program}
\vspace{-2ex}
  \end{figure}

  \begin{figure}[t]
    \scriptsize
    \centering
      $
      \begin{array}{l@{~}r@{~}l@{~}l@{~}l@{~}l@{~}l@{~}l@{~}l@{~}l@{~}}
         & \nil; & \tuple{\mathrm{c1},(\id,\underline{\mathsf{apply}~main/0~()}),\nil} \\[0ex]

        \longmapsto & \ldots \\[0ex]

        \longmapsto &
      \nil; & \tuple{\mathrm{c1},(\sigma,\mathsf{let}~\_=\underline{S\:!\: \{\mathrm{c1},req\}}~\mathsf{in}\ldots),\nil}
        ~ \&~\tuple{\mathrm{s},(\id,\mathsf{receive}~\{P,M\}\to\ldots),\nil}
        ~ \&\\
        &&\tuple{\mathrm{c2},(\sigma,ok),\nil}\\[0ex]

        \longmapsto &
        [(\mathrm{c1},\mathrm{s},[v_2])]; & \tuple{\mathrm{c1},(\sigma,\underline{\mathsf{let}~\_=\{\mathrm{c1},req\}~\mathsf{in}\ldots}),\nil}
        ~ \&~\tuple{\mathrm{s},(\id,\mathsf{receive}~\{P,M\}\to\ldots),\nil}
        ~ \&\\
        &&\tuple{\mathrm{c2},(\sigma,ok),\nil}\\[0ex]

        \longmapsto &
        [\underline{(\mathrm{c1},\mathrm{s},[v_2])}]; & \tuple{\mathrm{c1},(\sigma,\mathsf{receive}~ack\to\ldots),\nil}
        ~ \&~\tuple{\mathrm{s},(\id,\mathsf{receive}~\{P,M\}\to\ldots),\nil}
        ~ \&\\
        &&\tuple{\mathrm{c2},(\sigma,ok),\nil}\\[0ex]

        \longmapsto &
       \nil; & \tuple{\mathrm{c1},(\sigma,\mathsf{receive}~ack\to\ldots),\nil}
        ~ \&~\tuple{\mathrm{s},(\id,\underline{\mathsf{receive}~\{P,M\}\to\ldots}),[\{\mathrm{c1},req\}]}
        ~ \&\\
        &&\tuple{\mathrm{c2},(\sigma,ok),\nil}\\[0ex]

        \longmapsto &
       \nil; & \tuple{\mathrm{c1},(\sigma,\mathsf{receive}~ack\to\ldots),\nil}
        ~ \&~\tuple{\mathrm{s},(\theta_2,\mathsf{let}~\_=\underline{P\:!\: ack}~\mathsf{in}~\ldots),[\{\mathrm{c1},req\}]}
        ~ \&\\
        &&\tuple{\mathrm{c2},(\sigma,ok),\nil}\\[0ex]

        \longmapsto &
      [(\mathrm{s},\mathrm{c1},[ack])]; & \tuple{\mathrm{c1},(\sigma,\mathsf{receive}~ack\to\ldots),\nil}
        ~ \&~\tuple{\mathrm{s},(\theta_2,\underline{\mathsf{let}~\_=ack~\mathsf{in}~\ldots}),[\{\mathrm{c1},req\}]}
        ~ \&\\
        &&\tuple{\mathrm{c2},(\sigma,ok),\nil}\\[0ex]

        \longmapsto &
      [(\mathrm{s},\mathrm{c1},[ack])]; & \tuple{\mathrm{c1},(\sigma,\mathsf{receive}~ack\to\ldots),\nil}
        ~ \&~\tuple{\mathrm{s},(\theta_2,\underline{\mathsf{apply}~server/0~()}),\nil}
        ~ \&\\
        &&\tuple{\mathrm{c2},(\sigma,ok),\nil}\\[0ex]

        \longmapsto &
      [\underline{(\mathrm{s},\mathrm{c1},[ack])}]; &  \tuple{\mathrm{c1},(\sigma,\mathsf{receive}~ack\to\ldots),\nil}
        ~ \&~\tuple{\mathrm{s},(\id,\mathsf{receive}~\{P,M\}\to\ldots),\nil}
        ~ \&\\
        &&\tuple{\mathrm{c2},(\sigma,ok),\nil}\\[0ex]

        \longmapsto &
      \nil; & \tuple{\mathrm{c1},(\sigma,\underline{\mathsf{receive}~ack\to\ldots}),[ack]}
        ~ \&~\tuple{\mathrm{s},(\id,\mathsf{receive}~\{P,M\}\to\ldots),\nil}
        ~ \&\\
        &&\tuple{\mathrm{c2},(\sigma,ok),\nil}\\[0ex]

        \longmapsto &
      \nil; & \tuple{\mathrm{c1},(\sigma,ok),\nil}
        ~ \&~\tuple{\mathrm{s},(\id,\mathsf{receive}~\{P,M\}\to\ldots),\nil}
        ~ \&\\
        &&\tuple{\mathrm{c2},(\sigma,ok),\nil}\\

      \end{array}
      $
    \caption{A trace with $\sigma=\{S\mapsto \mathrm{s}\}$,
      $\theta_2 = \{P\mapsto \mathrm{c1},M\mapsto req\}$, and $v_2 =
      \{\mathrm{c1},req\}$.} \label{fig:ex2-derivation}
\vspace{-2ex}
  \end{figure}

\end{example}

\section{Reversible Semantics}

In this section, we introduce a reversible extension of the semantics
defined so far. 
Moreover, thanks to our modular design, the semantics for the language
expressions needs not be changed.

To be precise, in this section we introduce two transition relations:
$\rightharpoonup$ and $\leftharpoondown$. The first relation,
$\rightharpoonup$, is a conservative extension of the standard
semantics $\longmapsto$ (Figure~\ref{fig:system-rules}) to also
include some additional information in the states, following a typical
Landauer's embedding. We refer to $\rightharpoonup$ as the
\emph{forward} reversible semantics (or simply the forward
semantics). In contrast, the second relation, $\leftharpoondown$, proceeds in the backwards
direction, ``undoing'' the actions of a single process (and, by causal
consistency, possibly propagating it to other processes). We refer to
$\leftharpoondown$ as the backward (reversible) semantics.  Finally,
we denote the union $\rightharpoonup\cup\leftharpoondown$ by
$\rightleftharpoons$. Then, in a computation modelled with
$\rightleftharpoons$ the system mainly evolves forwards, except for
some processes that can run backwards in order to undo some particular
actions (and, afterwards, will run forwards again).

Here, we will introduce a (non-deterministic) ``undo'' operation (cf.\
rules \textit{Undo1} and \textit{Undo2} in Figure~\ref{marking-rules})
which has some similarities to, e.g., the rollback operator of
\cite{GLM14}. In order to delimit the scope of this operation (i.e.,
to determine when to stop undoing the actions of a process), we allow
the programmer to introduce \emph{checkpoints} in a
program. Syntactically, they are denoted with the built-in function
\textsf{check}, which is used in a let expression as follows: ``$
\mathsf{let}~\_=\mathsf{check}~\mathsf{in}~expr $''\!\!.  Function
$\mathsf{check}$ returns an identifier $\mathtt{t}$ associated to the
checkpoint (see below).
In the following, we consider that the rules to evaluate the language
expressions (Figures~\ref{fig:seq-rules} and
\ref{fig:concurrent-rules}) are extended with the following rule:
\[
  \hspace{18ex}(\mathit{Check}) ~~ {\displaystyle \frac{}{\theta,\mathsf{{check}
      \stackrel{\mathsf{check}(y)}{\too} \theta,y}}}
\]
In the next section, we will see that the only effect of a call to
function \textsf{check} is to add a checkpoint in the trace of a given
process.

\subsection{Forward Semantics}

First, we introduce the forward (reversible) semantics. Since the
expression rules are the same (except for the additional rule for
$\mathsf{check}$ mentioned above), we will only introduce the
reversible system rules, which are shown in
Figure~\ref{fig:system-rules-reversible}.  Processes now include a
memory $\pi$ that records the intermediate states of a process. Note
that we could optimize this in order to follow a strategy similar to
that in \cite{MHNHT07,NPV16,TA15} for the reversibility of functional
expressions, but this is orthogonal to our purpose in this paper, so
we focus only on the concurrent actions. 

The rules are mostly self-explanatory. Here, we use a special symbol
$\#_{k}^\mathtt{t}$ for checkpoints, where $k$ denotes the type of
checkpoint and $\mathtt{t}$ is a (unique) identifier for the
checkpoint. The checkpoints introduced by the programmer, denoted by
$\#_\mathsf{ch}^\mathtt{t}$, represent a \emph{safe} point in the
program execution, i.e., if we are undoing the actions of a given
process, we can safely stop the backward computation at any
checkpoint. Rollback operations (i.e., rules \textit{Undo1} and
\textit{Undo2} in Figure~\ref{marking-rules}) and checkpoints
introduced by the programmer lay the ground for defining \emph{safe}
sessions whose actions can be undone if, e.g., an exception occurs
before they are completed.
In practice, we distinguish three types of checkpoints:
\begin{itemize}
\item Checkpoints introduced by the programmer. These checkpoints
  correspond to the evaluation of function \textsf{check} and are
  denoted with $\#_\mathsf{ch}$.
\item Checkpoints associated to 
  receiving a message ($\#_\alpha$) and spawning a process
  ($\#_\mathsf{sp}$). These checkpoints are internal and only used to
  ensure causal consistency.
\end{itemize}
Also, we now assume that all messages $v$ are transformed into a tuple
$\{\mathtt{t},v\}$ where $\mathtt{t}$ is a unique identifier for this
message. Consequently, receive expressions should ignore this part of
the message when matching with the corresponding patterns.

\begin{figure}[t]
 \centering
  $
  \begin{array}{r@{~~}c}
    (\mathit{Internal}) & {\displaystyle
      \frac{\theta,e\stackrel{\tau}{\to} \theta',e'
      }{\Gamma;\tuple{\pi,\mathrm{p},(\theta,e),q}\& \Pi \rightharpoonup
      \Gamma;\tuple{\tau(\theta,e)\cons\pi,\mathrm{p},(\theta',e'),q}\& \Pi}
      }\\[4ex]

    (\mathit{Check}) & {\displaystyle
      \frac{\theta,e \stackrel{\mathsf{check}(y)}{\to} \theta',e'
                       ~~~\mbox{and $\mathtt{t}$ is fresh}}{\Gamma;\tuple{\pi,\mathrm{p},(\theta,e),q} 
        \& \Pi \rightharpoonup \Gamma;\tuple{\mathsf{check}(\theta,e)\cons\#_{\mathsf{ch}}^\mathtt{t}\cons\pi,\mathrm{p},(\theta',e'\{y\mapsto \mathtt{t}\}),q} 
        \& \Pi }
      }\\[4ex]

      (\mathit{Receive}) & {\displaystyle
        \frac{\theta,e \stackrel{\mathsf{rec}(y,\ol{cl_n})}{\to}
          \theta',e'~~~ \mathsf{matchrec}(\ol{cl_n},q) = (\theta_i,e_i,q',\mathtt{t})}{\Gamma;\tuple{\pi,\mathrm{p},(\theta,e),q}\& \Pi \rightharpoonup
          \Gamma;\tuple{\mathsf{rec}(\theta,e,q)\cons\pi,\mathrm{p},(\theta'\theta_i,e'\{y\mapsto e_i\}),q'}\& \Pi}
      }\\[4ex]
      
    (\mathit{Send}) & {\displaystyle
      \frac{\theta,e \stackrel{\mathsf{send}(\mathrm{p''},v)}{\to}
        \theta',e'~~~\mbox{and $\mathtt{t}$ is fresh}}{\Gamma;\tuple{\pi,\mathrm{p},(\theta,e),q} 
        \& \Pi \rightharpoonup
           \Gamma\cup (\mathrm{p},\mathrm{p''},\{\mathtt{t},v\});\tuple{\mathsf{send}(\mathrm{p''},\theta,e,\mathtt{t})\cons
           \pi,\mathrm{p},(\theta',e'),q}\& \Pi}
      }\\[4ex]

    (\mathit{Spawn}) & {\displaystyle
      \frac{\theta,e \stackrel{\mathsf{spawn}(y,a/n,[e_1,\ldots,e_n])}{\to}
        \theta',e'~~~ \mathrm{p'}~\mbox{is a fresh pid}~~~\mbox{and $\mathtt{t}$ is fresh}}{\begin{array}{ll}
          \Gamma;\tuple{\pi,\mathrm{p},(\theta,e),q} 
        \& \Pi \rightharpoonup &
        \Gamma;\tuple{\mathsf{spawn}(\theta,e,\mathrm{p'})\cons\pi,\mathrm{p},(\theta',e'\{y\mapsto
          \mathrm{p'}\}),q} \\
        & \& \tuple{\nil,\mathrm{p'},(\theta,(\mathsf{apply}~a/n~(e_1,\ldots,e_n)),\nil} 
        \& \Pi
      \end{array}}
      }\\[6ex]

    (\mathit{Self}) & {\displaystyle
      \frac{\theta,e \stackrel{\mathsf{self}(y)}{\to} \theta',e'}{\Gamma;\tuple{\pi,\mathrm{p},(\theta,e),q} 
        \& \Pi \rightharpoonup \Gamma;\tuple{\mathsf{self}(\theta,e)\cons\pi,\mathrm{p},(\theta',e'\{y\mapsto \mathrm{p}\}),q} 
        \& \Pi }
      }\\[4ex]

    (\mathit{Sched}) & {\displaystyle
      \frac{\alpha(\Gamma)=(\mathrm{p'},\mathrm{p})~~~~\Pi=\tuple{\pi,\mathrm{p},(\theta,e),q}\&\Pi'}{\Gamma;\Pi 
          \rightharpoonup
          \Gamma\setminus (\mathrm{p'},\mathrm{p},\{\mathtt{t},v\});\tuple{\alpha(\mathrm{p'},\mathrm{p},\{\mathtt{t},v\})\cons\pi,\mathrm{p},(\theta,e),\{\mathtt{t},v\}\cons q}\&\Pi' }
      }

  \end{array}
  $
\caption{Reversible semantics: system rules} \label{fig:system-rules-reversible}
\vspace{-2ex}
\end{figure}

\begin{example} \label{ex3} Consider again the program shown in
  Figure~\ref{fig:ex2-program}, where the function $client/1$ is now
  defined as follows:
  \[
  \begin{array}{r@{~}l@{~}l@{~}}
    client/1 = \mathsf{fun}~(S)\to &
    \mathsf{let}~\_=\mathsf{check}~\mathsf{in} & \mathsf{let}~\_=S\:!\:\{\mathsf{self}(),req\} \\
    && \mathsf{in}~\mathsf{receive}~\mathsf{ack}\to \mathsf{ok}~\mathsf{end} \\
  \end{array}
  \]
  and the execution trace shown in Figure~\ref{fig:ex2-derivation}.
  The corresponding forward (reversible) computation is
  shown in Figure~\ref{fig:ex3-derivation}. 

    \begin{figure}[t]
      \scriptsize
    \centering
      $
      \begin{array}{l@{~}r@{~}l@{~}l@{~}l@{~}l@{~}l@{~}l@{~}l@{~}l@{~}}
         & \nil; & \tuple{\nil,\mathrm{c1},(\id,\underline{\mathsf{apply}~main/0~()}),\nil} \\[0ex]

        \rightharpoonup & \ldots \\[0ex]

        \rightharpoonup &
      \nil; & \tuple{\pi_i,\mathrm{c1},(\sigma,\mathsf{let}~\_=\underline{\mathsf{check}}~\mathsf{in}\ldots),\nil}
        ~ \&~ \tuple{\pi'_i,\mathrm{s},(\id,\mathsf{receive}~\{P,M\}\to\ldots),\nil}
        ~ \&\\
        &&\tuple{\pi''_i,\mathrm{c2},(\sigma,ok),\nil}\\[.5ex]

        \rightharpoonup &
      \nil; & \tuple{\mathsf{check}(\sigma,\mathsf{let}~\_={\mathsf{check}}~\mathsf{in}\ldots)\cons\#_\mathsf{ch}^\mathsf{t_1}\cons\pi_i,\mathrm{c1},(\sigma,\underline{\mathsf{let}~\_=\mathsf{t_1}~\mathsf{in}\ldots}),\nil}
        ~ \&\\
        && \tuple{\pi'_i,\mathrm{s},(\id,\mathsf{receive}~\{P,M\}\to\ldots),\nil}
        ~ \&~\tuple{\pi''_i,\mathrm{c2},(\sigma,ok),\nil}\\[.5ex]

        \rightharpoonup &
      \nil; &
              \tuple{\tau(\sigma,\mathsf{let}~\_=\mathsf{t_1}~\mathsf{in}\ldots)\cons \mathsf{check}(\sigma,\mathsf{let}~\_={\mathsf{check}}~\mathsf{in}\ldots)\cons\#_\mathsf{ch}^\mathsf{t_1}\cons\pi_i,\\
        && ~\mathrm{c1},(\sigma,\mathsf{let}~\_=\underline{S\:!\: \{\mathrm{c1},req\}}~\mathsf{in}\ldots),\nil}
        ~ \&\\
        & & \tuple{\pi'_i,\mathrm{s},(\id,\mathsf{receive}~\{P,M\}\to\ldots),\nil}
        ~ \&~\tuple{\pi''_i,\mathrm{c2},(\sigma,ok),\nil}\\[.5ex]

        \rightharpoonup &
        [(\mathrm{c1},\mathrm{s},[\{\mathtt{t}_2,v_2\}])]; 
        &
          \tuple{\mathsf{send}(\mathrm{s},\sigma,\mathsf{let}~\_={S\:!\:
          \{\mathrm{c1},req\}}~\mathsf{in}\ldots,\mathtt{t}_2)\cons
          \tau(\sigma,\mathsf{let}~\_=\mathsf{t_1}~\mathsf{in}\ldots)\\
        &&~\cons \mathsf{check}(\sigma,\mathsf{let}~\_={\mathsf{check}}~\mathsf{in}\ldots)\cons\#_\mathsf{ch}^\mathsf{t_1}\cons\pi_i,\mathrm{c1},(\sigma,\underline{\mathsf{let}~\_=\{\mathrm{c1},req\}~\mathsf{in}\ldots}),\nil}
        ~ \&~\\
        && \tuple{\pi'_i,\mathrm{s},(\id,\mathsf{receive}~\{P,M\}\to\ldots),\nil}
        ~ \&~\tuple{\pi''_i,\mathrm{c2},(\sigma,ok),\nil}\\[.5ex]

        \rightharpoonup &
        [\underline{(\mathrm{c1},\mathrm{s},[\{\mathsf{t_2},v_2\}])}]; 
        & \tuple{\tau(\sigma,\mathsf{let}~\_=\{\mathrm{c1},req\}~\mathsf{in}\ldots)\cons \mathsf{send}(\mathrm{s},\sigma,\mathsf{let}~\_={S\:!\:
          \{\mathrm{c1},req\}}~\mathsf{in}\ldots,\mathtt{t}_2)\cons
          \tau(\sigma,\mathsf{let}~\_=\mathsf{t_1}~\mathsf{in}\ldots)\\
        &&~\cons \mathsf{check}(\sigma,\mathsf{let}~\_={\mathsf{check}}~\mathsf{in}\ldots)\cons\#_\mathsf{ch}^\mathsf{t_1}\cons\pi_i,\mathrm{c1},(\sigma,\mathsf{receive}~ack\to\ldots),\nil}
        ~ \&~\\
        && \tuple{\pi'_i,\mathrm{s},(\id,\mathsf{receive}~\{P,M\}\to\ldots),\nil}
        ~ \&~\tuple{\pi''_i,\mathrm{c2},(\sigma,ok),\nil}\\[.5ex]

        \rightharpoonup &
       \nil; & \tuple{\tau(\sigma,\mathsf{let}~\_=\{\mathrm{c1},req\}~\mathsf{in}\ldots)\cons \mathsf{send}(\mathrm{s},\sigma,\mathsf{let}~\_={S\:!\:
          \{\mathrm{c1},req\}}~\mathsf{in}\ldots,\mathtt{t}_2)\cons
          \tau(\sigma,\mathsf{let}~\_=\mathsf{t_1}~\mathsf{in}\ldots)\\
        &&~\cons \mathsf{check}(\sigma,\mathsf{let}~\_={\mathsf{check}}~\mathsf{in}\ldots)\cons\#_\mathsf{ch}^\mathsf{t_1}\cons\pi_i,\mathrm{c1},(\sigma,\mathsf{receive}~ack\to\ldots),\nil}
        ~ \&~\\
        &&\tuple{\alpha(\mathrm{c1},\mathrm{s},\{\mathsf{t_2},v_2\})\cons\pi'_i,\mathrm{s},(\id,\underline{\mathsf{receive}~\{P,M\}\to\ldots}),[\{\mathtt{t}_2,\{\mathrm{c1},req\}\}]}
        ~ \&\\
        &&\tuple{\pi''_i,\mathrm{c2},(\sigma,ok),\nil}\\[.5ex]

        \rightharpoonup &
       \nil; & \tuple{\tau(\sigma,\mathsf{let}~\_=\{\mathrm{c1},req\}~\mathsf{in}\ldots)\cons \mathsf{send}(\mathrm{s},\sigma,\mathsf{let}~\_={S\:!\:
          \{\mathrm{c1},req\}}~\mathsf{in}\ldots,\mathtt{t}_2)\cons
          \tau(\sigma,\mathsf{let}~\_=\mathsf{t_1}~\mathsf{in}\ldots)\\
        &&~\cons \mathsf{check}(\sigma,\mathsf{let}~\_={\mathsf{check}}~\mathsf{in}\ldots)\cons\#_\mathsf{ch}^\mathsf{t_1}\cons\pi_i,\mathrm{c1},(\sigma,\mathsf{receive}~ack\to\ldots),\nil}
        ~ \&~\\
        &&
           \tuple{\mathsf{rec}(\id,\mathsf{receive}~\{P,M\}\to\ldots,[\{\mathtt{t}_2,\{\mathrm{c1},req\}\}])\cons \alpha(\mathrm{c1},\mathrm{s},\{\mathsf{t_2},v_2\})\cons\pi'_i,\\
        && ~\mathrm{s},(\theta_2,\mathsf{let}~\_=\underline{P\:!\: ack}~\mathsf{in}~\ldots),\nil}
        ~ \&~ \tuple{\pi''_i,\mathrm{c2},(\sigma,ok),\nil}\\[0ex]


        \rightharpoonup & \ldots \\
      \end{array}
      $
    \caption{A (partial) trace with the forward reversible semantics.
    } \label{fig:ex3-derivation}
\vspace{-2ex}
  \end{figure}

\end{example}

\subsection{Backward Semantics}

Now, we present the backward semantics. The rules can be split into
two groups: the ones required to trigger, manage, and finish a rollback
operation, and the ones that actually perform the backward computation.

We denote a process running backwards with
$\lfloor\mathrm{p}\rfloor_\Psi$, where $\Psi$ is the set of
\emph{pending} checkpoints that the backward computation of
$\mathrm{p}$ has to go through before resuming its forward
computation.
For instance, a process of the form
$\lfloor\mathrm{p}\rfloor_{\{\#_\mathsf{ch}^\mathtt{t}\}}$ should go
backwards until a checkpoint $\#_\mathsf{ch}^\mathtt{t}$ is found in
its trace, a process
$\lfloor\mathrm{p}\rfloor_{\{\#_\alpha^\mathtt{t}\}}$ should go
backwards until an event of the form
$\alpha(\ldots,\{\mathtt{t},\_\})$ is found in its trace, and a
process $\lfloor\mathrm{p}\rfloor_{\{\#_\mathsf{sp}^\mathtt{t}\}}$
should go backwards until its initial state is reached (i.e., it
should be completely undone). Furthermore, a process may have a set
with several pending checkpoints coming from different rollback
operations.
Once a checkpoint is reached, it is removed from set $\Psi$. When
$\Psi = \emptyset$, the process $\mathrm{p}$ can resume its forward
computation (from the last checkpoint).

\begin{figure}[t]
 \centering
  \[
  \begin{array}{r@{~~}c}

    (\mathit{Undo1}) & {\displaystyle
      \Gamma;\tuple{\pi,\mathrm{p},(\theta,e),q} \:\&\: \Pi
        \leftharpoondown
      \Gamma;\lfloor \tuple{\pi,\mathrm{p},(\theta,e),q}\rfloor_{\#_\mathsf{ch}^\mathtt{t}} \:\&\: \Pi }
      \\[2ex]

    (\mathit{Undo2}) & {\displaystyle
      \Gamma;\lfloor\tuple{\pi,\mathrm{p},(\theta,e),q}\rfloor_{\Psi} \:\&\: \Pi
        \leftharpoondown
      \Gamma;\lfloor
      \tuple{\pi,\mathrm{p},(\theta,e),q}\rfloor_{\Psi\cup \#_\mathsf{ch}^\mathtt{t}}\: \&\: \Pi }
      \\[2ex]

    (\mathit{Stop1}) & {\displaystyle
      \Gamma;\lfloor\tuple{\pi,\mathrm{p},(\theta,e),q}\rfloor_{\emptyset} \:\&\: \Pi
        \leftharpoondown
      \Gamma;\tuple{\pi,\mathrm{p},(\theta,e),q}\: \&\: \Pi }
      \\[2ex]

    (\mathit{Stop2}) & {\displaystyle
      \Gamma;\lfloor\tuple{\nil,\mathrm{p},(\theta,e),q}\rfloor_{\{\#_\mathsf{sp}^\mathtt{t}\}} \:\&\: \Pi
        \leftharpoondown
      \Gamma;\Pi }
      \\[2ex]

    (\mathit{Check}) & {\displaystyle
      \Gamma;\lfloor\tuple{\#_\mathsf{ch}^\mathtt{t}\cons\pi,\mathrm{p},(\theta,e),q}\rfloor_{\Psi\cup \#_\mathsf{ch}^\mathtt{t}} \:\&\: \Pi
        \leftharpoondown
      \Gamma;\lfloor \tuple{\pi,\mathrm{p},(\theta,e),q}\rfloor_{\Psi}\: \&\: \Pi }
      \\[2ex]

    (\mathit{Discard}) & {\displaystyle
       \Gamma;\lfloor\tuple{\#_\mathsf{ch}^\mathtt{t}\cons\pi,\mathrm{p},(\theta,e),q}\rfloor_{\Psi} 
        \:\&\: \Pi \leftharpoondown
      \Gamma;\lfloor\tuple{\pi,\mathrm{p},(\theta,e),q}\rfloor_{\Psi} 
        \:\&\: \Pi  ~~~~
         \mbox{if $\#_\mathsf{ch}^\mathtt{t}\not\in\Psi$}
      } 

  \end{array}
  \]
\caption{Backward semantics: rules for dealing with rollbacks}
\label{marking-rules}
\vspace{-2ex}
\end{figure}

Let us now explain the rules shown in Figure~\ref{marking-rules},
whose purpose is to mark (or unmark) a process for backward
computation.
The first two rules, $\mathit{Undo1}$ and $\mathit{Undo2}$, are used
to introduce a new rollback for a checkpoint $\#_\mathsf{ch}^\mathtt{t}$
for some $\mathtt{t}$.  

A backward computation ends when there are no pending checkpoints in
the set (rule $\mathit{Stop1}$) or when we reach a process with an
empty trace and a spawn checkpoint $\#_\mathsf{sp}^\mathtt{t}$ in the
set of pending checkpoints, so that the process is removed from the
system (rule $\mathit{Stop2}$).

Finally, if we reach a checkpoint $\#_\mathsf{ch}^\mathtt{t}$ in the
trace of a process running backwards, we can either remove it from the
set of pending checkpoints (rule $\mathit{Check}$) or just ignore it
when it was not in the set of pending checkpoints (rule
$\mathit{Discard}$).\footnote{Note that this may happen often since
  the evaluation of each function \textsf{check} introduces a
  checkpoint $\#_\mathsf{ch}^\mathtt{t}$ no matter whether a rollback
  operation is run or not.}


\begin{figure}[hbt]
 \centering
  \[
  \begin{array}{r@{~~}c}
      (\mathit{Internal}) & {\displaystyle
        \Gamma;\lfloor
                            \tuple{\tau(\theta,e)\cons\pi,p,(\theta',e'),q}\rfloor_{\Psi}
                            \:\&\: \Pi
          \leftharpoondown  \Gamma;\lfloor\tuple{\pi,p,(\theta,e),q}\rfloor_{\Psi}\:\&\: \Pi
        }
      \\[2ex]

      (\mathit{Check}) & {\displaystyle
        \Gamma;\lfloor
                            \tuple{\mathsf{check}(\theta,e)\cons\pi,p,(\theta',e'),q}\rfloor_{\Psi}
                            \:\&\: \Pi
          \leftharpoondown  \Gamma;\lfloor\tuple{\pi,p,(\theta,e),q}\rfloor_{\Psi}\:\&\: \Pi
        }
      \\[2ex]

      (\mathit{Receive}) & {\displaystyle
        \Gamma;\lfloor\tuple{\mathsf{rec}(\theta,e,q)\cons\pi,\mathrm{p},(\theta',e'),q'}\rfloor_{\Psi}\:\&\: \Pi
          \leftharpoondown  \Gamma;\lfloor\tuple{\pi,\mathrm{p},(\theta,e),q}\rfloor_{\Psi}\:\&\: \Pi
        }
      \\[2ex]
      
    (\mathit{Send1}) & {\displaystyle
      \begin{array}{l}
        \Gamma;\lfloor\tuple{\mathsf{send}(\mathrm{p''},\theta,e,\mathtt{t})\cons\pi,\mathrm{p},(\theta',e'),q}\rfloor_{\Psi}\:\&\:
        \Pi
        \leftharpoondown \Gamma';\lfloor\tuple{\pi,\mathrm{p},(\theta,e),q}\rfloor_{\Psi} 
        \:\&\: \Pi\\
      \hspace{20ex}\mbox{if}~(\mathrm{p},\mathrm{p''},\{\mathtt{t},v\})~\mbox{occurs
          in}~\Gamma,~\mbox{with}~\Gamma' =
        \Gamma\setminus\!\!\!\setminus(\mathrm{p},\mathrm{p''},v,\mathtt{t})
    \end{array}
  }\\[3ex]

    (\mathit{Send2}) & {\displaystyle
      \begin{array}{l}
        \Gamma;\lfloor\tuple{\mathsf{send}(\mathrm{p''},\theta,e,\mathtt{t})\cons\pi,\mathrm{p},(\theta',e'),q}\rfloor_{\Psi}\:\&\:
        \tuple{\pi'',\mathrm{p''},(\theta'',e''),q''}\:\&\: \Pi\\
        \hspace{15ex}\leftharpoondown
        \Gamma;\lfloor\tuple{\pi,\mathrm{p},(\theta,e),q}\rfloor_{\Psi} \:\&\:
        \lfloor\tuple{\pi'',\mathrm{p''},(\theta'',e''),q''}\rfloor_{\#_\alpha^\mathtt{t}}\:\&\:
        \Pi \\
      \hspace{20ex}\mbox{if}~(\mathrm{p},\mathrm{p''},\{\mathtt{t},v\})~\mbox{does
        not occur in}~\Gamma
    \end{array}
  }\\[5ex]

    (\mathit{Spawn}) & {\displaystyle
      \begin{array}{l}
        \Gamma;\lfloor\tuple{\mathsf{spawn}(\theta,e,\mathrm{p''})\cons\pi,\mathrm{p},(\theta',e'),q}\rfloor_{\Psi}
        \:\&\: 
             \tuple{\pi'',\mathrm{p''},(\theta'',e''),q'')} 
        \:\&\: \Pi
        \\
        \hspace{10ex}\leftharpoondown
      \Gamma;\lfloor\tuple{\pi,\mathrm{p},(\theta,e),q}\rfloor_{\Psi}
        \:\&\: 
             \lfloor\tuple{\pi'',\mathrm{p''},(\theta'',e''),q'')}\rfloor_{\#_\mathsf{sp}^\mathrm{p''}}
        \:\&\: \Pi 
      \end{array}
      }\\[4ex]

    (\mathit{Self}) & {\displaystyle
       \Gamma;\lfloor\tuple{\mathsf{self}(\theta,e)\cons\pi,\mathrm{p},(\theta',e'),q}\rfloor_{\Psi} 
        \:\&\: \Pi \leftharpoondown
      \Gamma;\lfloor\tuple{\pi,\mathrm{p},(\theta,e),q}\rfloor_{\Psi} 
        \:\&\: \Pi  
      }\\[2ex]

    (\mathit{Sched1}) & {\displaystyle
      \begin{array}{l}
      \Gamma;\lfloor\tuple{\alpha(\mathrm{p''},\mathrm{p},\{\mathtt{t},v\})\cons\pi,\mathrm{p},(\theta,e),\{\mathtt{t},v\}\cons
        q}\rfloor_{\Psi\cup \#_\alpha^\mathtt{t}} \: \&\: \Pi 
\leftharpoondown 
      \Gamma;\lfloor\tuple{\pi,\mathrm{p},(\theta,e),q}\rfloor_{\Psi}\:\&\:\Pi 
    \end{array}
    }\\[2ex]

    (\mathit{Sched2}) & {\displaystyle
      \begin{array}{l}
      \Gamma;\lfloor\tuple{\alpha(\mathrm{p''},\mathrm{p},\{\mathtt{t},v\})\cons\pi,\mathrm{p},(\theta,e),\{\mathtt{t},v\}\cons
        q}\rfloor_{\Psi} \: \&\: \Pi \\ 
      \hspace{15ex}\leftharpoondown 
      \Gamma\:\ol{\cup}\:(\mathrm{p''},\mathrm{p},\{\mathtt{t},v\});\lfloor\tuple{\pi,\mathrm{p},(\theta,e),q}\rfloor_{\Psi}\:\&\:\Pi
      ~~~~   \mbox{if $\#_\mathsf{\alpha}^\mathtt{t}\not\in\Psi$}
    \end{array}
    }
  \end{array}
  \]
\caption{Backward semantics: Rules for backward computation.} \label{concretesem4}
\vspace{-2ex}
\end{figure}

Let us now discuss the rules for performing backward computations,
which are shown in Figure~\ref{concretesem4}.  In general, these rules
are applied whenever the term located at the top of the computation
history $\pi$ is an event associated to a sequential or concurrent
expression (i.e., when it is not a checkpoint). These terms are
deleted from the trace when any of these rules is applied.

In general, all rules restore the control (and sometimes also the
queue) of a process. 
In order to reverse the creation of a process, rule $\mathit{Spawn}$
marks the child process with checkpoint $\#_\mathsf{sp}^\mathrm{p'}$.
The child process will then run backwards until, eventually, its trace
is empty and rule $\mathit{Stop2}$ removes it from the system.

For undoing the sending of a message, rule $\mathit{Send1}$ removes a
message from $\Gamma$ when the message has not been delivered
yet. Here, we use the operator ``$\setminus\!\setminus$'' to denote
the removal of a message no matter its position (in contrast to
``$\setminus$'' which always removes the oldest message of the list,
i.e., the first one).
Otherwise, rule $\mathit{Send2}$ propagates the backward
computation to the receiver process by adding $\#_\alpha^\mathtt{t}$
as a pending checkpoint of the receiver process ($\mathtt{t}$
identifies the message). This will cause the receiver process to undo
all the actions that it has performed since it received the message,
thus ensuring \emph{causal consistency}. 

Observe that, at first glance, one may think rule $\mathit{Receive}$
should also introduce some new rollback operation for causal
consistency. However, if we take a closer look, we will realize that
receiving a message in our context is just about processing the
message, and not actually receiving it.  In fact, the processed
message could have been delivered to the process mailbox a long time
ago, and triggering a backward computation on the sending process
would be unnecessary.

It is actually rules $\mathit{Sched1}$ and $\mathit{Sched2}$ that take
care of dealing with terms of the form
$\alpha(\mathrm{p}'',\mathrm{p},\{\mathtt{t},v\})$ in a trace. When the
rollback was introduced by another process (in rule $\mathit{Send2}$),
rule $\mathit{Sched1}$ removes the corresponding message
$\{\mathtt{t},v\}$ from the process queue and deletes the checkpoint
$\#_\alpha^\mathtt{t}$ from the set of pending checkpoints. No further
action is required; actually, when the process $\mathrm{p}''$ that
triggered rule $\mathit{Send2}$ resumes its forward computation, the
same message $v$ (with a different identifier $\mathtt{t}'$) will be
delivered again to process $\mathrm{p}$.
When the rollback was not introduced using rule $\mathit{Send2}$,
i.e., there is no checkpoint $\#_\alpha^\mathtt{t}$ in $\Psi$, then
$\mathit{Sched2}$ proceeds as before but also adds the message
$(\mathrm{p}'',\mathrm{p},\{\mathtt{t},v\})$ to the global mailbox
$\Gamma$. Then, when this process resumes its forward computation, the
same message will again be delivered by the scheduler. Here, the
operator $\ol{\cup}$ is used to add messages to the head of the
corresponding list instead of to its end, i.e., $\Gamma\:\ol{\cup}\:
(\mathrm{p},\mathrm{q},\{\mathtt{t},v\})$ denotes
$\Gamma\setminus\{(\mathrm{p},\mathrm{q},vs)\} \cup
\{(\mathrm{p},\mathrm{q},\{\mathtt{t},v\}:vs)\}$.

In both cases, we aim at eventually producing a system that could have
been obtained from the initial state using only forward steps.

\begin{example} \label{ex4} Consider the forward execution trace shown
  in Figure~\ref{fig:ex3-derivation}.  A corresponding backward
  computation is shown in Figure~\ref{fig:ex4-derivation}. 

    \begin{figure}[t]
      \scriptsize
    \centering
      $
      \begin{array}{l@{~}r@{~}l@{~}l@{~}l@{~}l@{~}l@{~}l@{~}l@{~}l@{~}}

        & \nil; & \blue{\lfloor\langle \tau(\sigma,\mathsf{let}~\_=\{\mathrm{c1},req\}~\mathsf{in}\ldots)\cons \mathsf{send}(\mathrm{s},\sigma,\mathsf{let}~\_={S\:!\:
          \{\mathrm{c1},req\}}~\mathsf{in}\ldots,\mathtt{t}_2)\cons
          \tau(\sigma,\mathsf{let}~\_=\mathsf{t_1}~\mathsf{in}\ldots)}\\
        &&~\blue{\cons \mathsf{check}(\sigma,\mathsf{let}~\_={\mathsf{check}}~\mathsf{in}\ldots)\cons\#_\mathsf{ch}^\mathsf{t_1}\cons\pi_i,\mathrm{c1},(\sigma,\mathsf{receive}~ack\to\ldots),\nil\rangle\rfloor_{\#_{\mathsf{ch}}^{\mathtt{t}_1}}}
        ~ \&~\\
        &&  \tuple{\mathsf{rec}(\id,\mathsf{receive}~\{P,M\}\to\ldots,[\{\mathtt{t}_2,\{\mathrm{c1},req\}\}])\cons \alpha(\mathrm{c1},\mathrm{s},\{\mathsf{t_2},v_2\})\cons\pi'_i,\\
        && ~\mathrm{s},(\theta_2,\mathsf{let}~\_={P\:!\: ack}~\mathsf{in}~\ldots),\nil}
        ~ \&~ \tuple{\pi''_i,\mathrm{c2},(\sigma,ok),\nil}\\[.5ex]

        \leftharpoondown
        & \nil; & \blue{\lfloor\langle \mathsf{send}(\mathrm{s},\sigma,\mathsf{let}~\_={S\:!\:
          \{\mathrm{c1},req\}}~\mathsf{in}\ldots,\mathtt{t}_2)\cons
          \tau(\sigma,\mathsf{let}~\_=\mathsf{t_1}~\mathsf{in}\ldots)}\\
        &&~\blue{\cons \mathsf{check}(\sigma,\mathsf{let}~\_={\mathsf{check}}~\mathsf{in}\ldots)\cons\#_\mathsf{ch}^\mathsf{t_1}\cons\pi_i,\mathrm{c1},(\sigma,\mathsf{let}~\_=\{\mathrm{c1},req\}~\mathsf{in}\ldots),\nil\rangle\rfloor_{\#_{\mathsf{ch}}^{\mathtt{t}_1}}}
        ~ \&~\\
        &&  \tuple{\mathsf{rec}(\id,\mathsf{receive}~\{P,M\}\to\ldots,[\{\mathtt{t}_2,\{\mathrm{c1},req\}\}])\cons \alpha(\mathrm{c1},\mathrm{s},\{\mathsf{t_2},v_2\})\cons\pi'_i,\\
        && ~\mathrm{s},(\theta_2,\mathsf{let}~\_={P\:!\: ack}~\mathsf{in}~\ldots),\nil}
        ~ \&~ \tuple{\pi''_i,\mathrm{c2},(\sigma,ok),\nil}\\[.5ex]


        \leftharpoondown
        & \nil; & \blue{\lfloor\langle 
          \tau(\sigma,\mathsf{let}~\_=\mathsf{t_1}~\mathsf{in}\ldots)}\\
        &&~\blue{\cons \mathsf{check}(\sigma,\mathsf{let}~\_={\mathsf{check}}~\mathsf{in}\ldots)\cons\#_\mathsf{ch}^\mathsf{t_1}\cons\pi_i,\mathrm{c1},(\sigma,\mathsf{let}~\_={S\:!\:
          \{\mathrm{c1},req\}}~\mathsf{in}\ldots,\mathtt{t}_2),\nil\rangle\rfloor_{\#_{\mathsf{ch}}^{\mathtt{t}_1}}}
        ~ \&~\\
        &&  \blue{\lfloor\langle\mathsf{rec}(\id,\mathsf{receive}~\{P,M\}\to\ldots,[\{\mathtt{t}_2,\{\mathrm{c1},req\}\}])\cons \alpha(\mathrm{c1},\mathrm{s},\{\mathsf{t_2},v_2\})\cons\pi'_i,}\\
        && ~\blue{\mathrm{s},(\theta_2,\mathsf{let}~\_={P\:!\: ack}~\mathsf{in}~\ldots),\nil\rangle\rfloor_{\#_\alpha^{\mathtt{t}_2}}}
        ~ \&~ \tuple{\pi''_i,\mathrm{c2},(\sigma,ok),\nil}\\[.5ex]

        \leftharpoondown
        & \nil; & \blue{\lfloor\langle 
          \mathsf{check}(\sigma,\mathsf{let}~\_={\mathsf{check}}~\mathsf{in}\ldots)\cons\#_\mathsf{ch}^\mathsf{t_1}\cons\pi_i,\mathrm{c1},(\sigma,\mathsf{let}~\_=\mathsf{t_1}~\mathsf{in}\ldots),\nil\rangle\rfloor_{\#_{\mathsf{ch}}^{\mathtt{t}_1}}}
        ~ \&~\\
        &&  \blue{\lfloor\langle\mathsf{rec}(\id,\mathsf{receive}~\{P,M\}\to\ldots,[\{\mathtt{t}_2,\{\mathrm{c1},req\}\}])\cons \alpha(\mathrm{c1},\mathrm{s},\{\mathsf{t_2},v_2\})\cons\pi'_i,}\\
        && ~\blue{\mathrm{s},(\theta_2,\mathsf{let}~\_={P\:!\: ack}~\mathsf{in}~\ldots),\nil\rangle\rfloor_{\#_\alpha^{\mathtt{t}_2}}}
        ~ \&~ \tuple{\pi''_i,\mathrm{c2},(\sigma,ok),\nil}\\[.5ex]

        \leftharpoondown
        & \nil; & \blue{\lfloor\langle 
          \#_\mathsf{ch}^\mathsf{t_1}\cons\pi_i,\mathrm{c1},(\sigma,\mathsf{let}~\_={\mathsf{check}}~\mathsf{in}\ldots),\nil\rangle\rfloor_{\#_{\mathsf{ch}}^{\mathtt{t}_1}}}
        ~ \&~\\
        &&  \blue{\lfloor\langle\mathsf{rec}(\id,\mathsf{receive}~\{P,M\}\to\ldots,[\{\mathtt{t}_2,\{\mathrm{c1},req\}\}])\cons \alpha(\mathrm{c1},\mathrm{s},\{\mathsf{t_2},v_2\})\cons\pi'_i,}\\
        && ~\blue{\mathrm{s},(\theta_2,\mathsf{let}~\_={P\:!\: ack}~\mathsf{in}~\ldots),\nil\rangle\rfloor_{\#_\alpha^{\mathtt{t}_2}}}
        ~ \&~ \tuple{\pi''_i,\mathrm{c2},(\sigma,ok),\nil}\\[.5ex]

        \leftharpoondown
        & \nil; & \blue{\lfloor\langle 
          \pi_i,\mathrm{c1},(\sigma,\mathsf{let}~\_={\mathsf{check}}~\mathsf{in}\ldots),\nil\rangle\rfloor_{\emptyset}}
        ~ \&~\\
        &&  \blue{\lfloor\langle\mathsf{rec}(\id,\mathsf{receive}~\{P,M\}\to\ldots,[\{\mathtt{t}_2,\{\mathrm{c1},req\}\}])\cons \alpha(\mathrm{c1},\mathrm{s},\{\mathsf{t_2},v_2\})\cons\pi'_i,}\\
        && ~\blue{\mathrm{s},(\theta_2,\mathsf{let}~\_={P\:!\: ack}~\mathsf{in}~\ldots),\nil\rangle\rfloor_{\#_\alpha^{\mathtt{t}_2}}}
        ~ \&~ \tuple{\pi''_i,\mathrm{c2},(\sigma,ok),\nil}\\[.5ex]

        \leftharpoondown
        & \nil; & \langle 
          \pi_i,\mathrm{c1},(\sigma,\mathsf{let}~\_={\mathsf{check}}~\mathsf{in}\ldots),\nil\rangle
        ~ \&~\\
        &&  \blue{\lfloor\langle\mathsf{rec}(\id,\mathsf{receive}~\{P,M\}\to\ldots,[\{\mathtt{t}_2,\{\mathrm{c1},req\}\}])\cons \alpha(\mathrm{c1},\mathrm{s},\{\mathsf{t_2},v_2\})\cons\pi'_i,}\\
        && ~\blue{\mathrm{s},(\theta_2,\mathsf{let}~\_={P\:!\: ack}~\mathsf{in}~\ldots),\nil\rangle\rfloor_{\#_\alpha^{\mathtt{t}_2}}}
        ~ \&~ \tuple{\pi''_i,\mathrm{c2},(\sigma,ok),\nil}\\[.5ex]

        \leftharpoondown
        & \nil; & \langle 
          \pi_i,\mathrm{c1},(\sigma,\mathsf{let}~\_={\mathsf{check}}~\mathsf{in}\ldots),\nil\rangle
        ~ \&~\\
        &&  \blue{\lfloor\langle\alpha(\mathrm{c1},\mathrm{s},\{\mathsf{t_2},v_2\})\cons\pi'_i,}
         ~\blue{\mathrm{s},(\id,\mathsf{receive}~\{P,M\}\to\ldots),[\{\mathtt{t}_2,\{\mathrm{c1},req\}\}]\rangle\rfloor_{\#_\alpha^{\mathtt{t}_2}}}
        ~ \&\\
        &&\tuple{\pi''_i,\mathrm{c2},(\sigma,ok),\nil}\\[.5ex]


        \leftharpoondown
        & \nil; & \langle 
          \pi_i,\mathrm{c1},(\sigma,\mathsf{let}~\_={\mathsf{check}}~\mathsf{in}\ldots),\nil\rangle
        ~ \&~\\
        &&  \blue{\lfloor\langle\pi'_i,\mathrm{s},(\id,\mathsf{receive}~\{P,M\}\to\ldots),\nil\rangle\rfloor_{\emptyset}}
        ~ \&~\tuple{\pi''_i,\mathrm{c2},(\sigma,ok),\nil}\\[.5ex]

        \leftharpoondown
        & \nil; & \langle 
          \pi_i,\mathrm{c1},(\sigma,\mathsf{let}~\_={\mathsf{check}}~\mathsf{in}\ldots),\nil\rangle
        ~ \&~\\
        &&  \langle\pi'_i,\mathrm{s},(\id,\mathsf{receive}~\{P,M\}\to\ldots),\nil\rangle
        ~ \&~\tuple{\pi''_i,\mathrm{c2},(\sigma,ok),\nil}\\

      \end{array}
      $
    \caption{A (partial) trace with the backward reversible semantics.
    } \label{fig:ex4-derivation}
 \vspace{-2ex}
 \end{figure}

\end{example}
The soundness of our reversible semantics can be stated as follows:
i) every forward step can be reversed;
ii) every system reached during a computation with
  $\rightleftharpoons$, could have been reached with
  $\rightharpoonup$ from the initial system.

\section{Discussion}

We have defined a reversible semantics for a first-order subset of
Erlang. To the best of our knowledge, this is the first attempt to
define a reversible semantics for Erlang.
As mentioned in the introduction, the closest to our work is the
debugging approach based on a rollback construct of
\cite{GLM14,LMSS11,LMS16,LLMS12}, but it is defined in the context of
a higher-order asynchronous $\pi$-calculus. Also, we share some
similarities with the checkpointing technique for fault-tolerant
distributed computing of \cite{FV05,KFV14}, although the aim is
different (they aim at defining a new language rather than extending
an existing one).

As future work, we plan to formally study the correctness of our
calculi, as claimed at the end of the previous section. Also, we would
like to extend the Erlang language with a new construct for \emph{safe
  sessions} so that all the actions in a session can be undone when
the session aborts, which has a great potential to automate the
fault-tolerance capabilities of the language Erlang.

\subsection*{Acknowledgements}

We would like to thank Ivan Lanese and the anonymous reviewers for
many useful suggestions to improve this paper.


\begin{thebibliography}{10}
\providecommand{\url}[1]{\texttt{#1}}
\providecommand{\urlprefix}{URL }

\bibitem{AVW96}
Armstrong, J., Virding, R., Williams, M.: {Concurrent programming in Erlang
  (2nd edition)}. Prentice Hall (1996)

\bibitem{Ben73}
Bennett, C.: Logical reversibility of computation. IBM Journal of Research and
  Development  17,  525--532 (1973)

\bibitem{Bennett00}
Bennett, C.: Notes on the history of reversible computation. {IBM} Journal of
  Research and Development  44(1),  270--278 (2000)

\bibitem{CMRT13tr}
Caballero, R., Mart{\'{\i}}n-Mart{\'{\i}}n, E., Riesco, A., Tamarit, S.: A
  declarative debugger for concurrent erlang programs (extended version). Tech.\
  Rep.\ SIC-15/13, UCM (2013),
  \url{\texttt{http://maude.sip.ucm.es/$\sim$adrian/files/conc\_cal\_tr.pdf}}

\bibitem{CL11}
Cardelli, L., Laneve, C.: Reversible structures. In Proc.\ of {CMSB} 2011, pp. 131--140. {ACM} (2011)

\bibitem{CGJLNPV04}
Carlsson, R., Gustavsson, B., Johansson, E. \emph{et al}: Core erlang 1.0.3. language specification
  (2004), available from \verb$https://www.it.uu.se/research/$
  \verb$group/hipe/cerl/doc/core_erlang-1.0.3.pdf$

\bibitem{DK04}
Danos, V., Krivine, J.: Reversible communicating systems. In Proc.\ of
{CONCUR} 2004. Springer LNCS 3170, pp.
  292--307 (2004)

\bibitem{DK05}
Danos, V., Krivine, J.: Transactions in {RCCS}. In Proc.\ of {CONCUR}
2005. Springer LNCS 3653, pp. 398--412 (2005)

\bibitem{FV05}
Field, J., Varela, C.A.: Transactors: a programming model for maintaining
  globally consistent distributed state in unreliable
  environments.  In Proc.\ of {POPL} 2005, pp. 195--208.
  {ACM} (2005)

\bibitem{Fra05}
Frank, M.P.: Introduction to reversible computing: motivation, progress, and
  challenges. In Proc.\ of 2nd Conf.\ on Computing Frontiers, pp. 385--390.
  {ACM} (2005)

\bibitem{GLM14}
Giachino, E., Lanese, I., Mezzina, C.A.: Causal-consistent reversible
  debugging. In Proc.\ of {FASE} 2014. Springer LNCS 8411, pp. 370--384 (2014)

\bibitem{KFV14}
Kuang, P., Field, J., Varela, C.A.: Fault tolerant distributed computing using
  asynchronous local checkpointing. In Proc.\ of AGERE! 2014,
  pp. 81--93. {ACM} (2014)

\bibitem{Lan61}
Landauer, R.: Irreversibility and heat generation in the computing process. IBM
  Journal of Research and Development  5,  183--191 (1961)

\bibitem{LMSS11}
Lanese, I., Mezzina, C.A., Schmitt, A., Stefani, J.: Controlling reversibility
  in higher-order pi. In Proc.\ {CONCUR} 2011. Springer LNCS 6901, pp. 297--311 (2011)

\bibitem{LMS16}
Lanese, I., Mezzina, C.A., Stefani, J.: Reversibility in the higher-order
  {\(\pi\)}-calculus. Theor. Comput. Sci.  625,  25--84 (2016)

\bibitem{LLMS12}
Lienhardt, M., Lanese, I., Mezzina, C.A., Stefani, J.: A reversible abstract
  machine and its space overhead. In Proc.\ of
  the Joint {FMOODS} 2012 and {FORTE} 2012 Int'l Conference. Springer
  LNCS 7273, pp. 1--17  (2012)

\bibitem{MHNHT07}
Matsuda, K., Hu, Z., Nakano, K., Hamana, M., Takeichi, M.: Bidirectionalization
  transformation based on automatic derivation of view complement functions.
  In Proc.\ of {ICFP} 2007, pp. 47--58.
  {ACM} (2007)

\bibitem{NPV16}
Nishida, N., Palacios, A., Vidal, G.: Reversible term rewriting. In: Kesner,
  D., Pientka, B. (eds.) Proc.\ of the First International Conference on Formal
  Structures for Computation and Deduction (FSCD'16). Leibniz International
  Proceedings in Informatics (2016)

\bibitem{PU07}
Phillips, I., Ulidowski, I.: Reversing algebraic process calculi. J. Log.
  Algebr. Program.  73(1-2),  70--96 (2007)

\bibitem{SFB10}
Svensson, H., Fredlund, L.A., Earle, C.B.: {A unified semantics for future
  Erlang}. In Proc.\ of the 9th ACM SIGPLAN workshop on Erlang, pp. 23--32.
  ACM (2010)

\bibitem{TA15}
Thomsen, M.K., Axelsen, H.B.: {Interpretation and programming of the reversible
  functional language RFUN}. In Proc.\ of IFL 2015. Springer
  (2016), \emph{to appear}

\bibitem{TY15}
Tiezzi, F., Yoshida, N.: Reversible session-based pi-calculus. J. Log. Algebr.
  Meth. Program.  84(5),  684--707 (2015)

\bibitem{Yok10}
Yokoyama, T.: Reversible computation and reversible programming languages.
  Electronic Notes in Theoretical Computer Science  253(6),  71--81 (2010),
  proc.\ of the Workshop on Reversible Computation (RC 2009)

\bibitem{YAG08}
Yokoyama, T., Axelsen, H.B., Gl{\"{u}}ck, R.: Principles of a reversible
  programming language. In Proc.\ of the 5th Conference on Computing Frontiers, pp. 43--54. {ACM}
  (2008)

\bibitem{YAG16}
Yokoyama, T., Axelsen, H.B., Gl{\"{u}}ck, R.: Fundamentals of reversible
  flowchart languages. Theor. Comput. Sci.  611,  87--115 (2016)

\end{thebibliography}
\end{document}